\documentclass{IEEEtran}
\usepackage{cite}
\usepackage{amsmath,amssymb,amsfonts}
\usepackage{algorithmic}
\usepackage{graphicx}
\usepackage{textcomp}
\begin{document}
\title{Generative Adversarial Networks in Digital Pathology: A Survey on Trends and Future Potential}
\author{M. E. Tschuchnig, \IEEEmembership{Member, IEEE}, G. J. Oostingh, and M. Gadermayr
\thanks{Manuscript submitted April 30, 2020. This work was partially funded by the County of Salzburg under grant number FHS-2019-10-KIAMed.}
\thanks{M. E. Tschuchnig is at the Department of Information Technologies and Systems Management, Salzburg University of Applied Sciences, Salzburg, Austria,
and also at the Department of Biomedical Sciences,
Salzburg University of Applied Sciences, Salzburg, Austria, (e-mail: maximilian.tschuchnig@fh-salzburg.ac.at).}
\thanks{G. J. Oostingh is at the Department of Biomedical Sciences, Salzburg University of Applied Sciences, Salzburg, Austria (e-mail: geja.oostingh@fh-salzburg.ac.at).}
\thanks{M. Gadermayr is at the Department of Information Technologies and Systems Management, Salzburg University of Applied Sciences, Salzburg, Austria (e-mail: michael.gadermayr@fh-salzburg.ac.at).}}

\maketitle







\begin{abstract}
	Image analysis in the field of digital pathology has recently gained increased popularity.
	The use of high-quality whole slide scanners enables the fast acquisition of large amounts of image data, showing extensive context and microscopic detail at the same time.
	Simultaneously, novel machine learning algorithms have boosted the performance of image analysis approaches.
	In this paper, we focus on a particularly powerful class of architectures, called Generative Adversarial Networks (GANs), applied to histological image data.
	Besides improving performance, GANs also enable application scenarios in this field, which were previously intractable.
	However, GANs could exhibit a potential for introducing bias.
	Hereby, we summarize the recent state-of-the-art developments in a generalizing notation, present the main applications of GANs and give an outlook of some chosen promising approaches and their possible future applications. In addition, we identify currently unavailable methods with potential for future applications.
\end{abstract}

\section{Motivation}

Whole slide scanners are capable of effectively digitizing histo- or cytological slides without any significant manual effort.
These scanners are capable of generating vast amounts of digital data, as even a single whole slide image can show up to several giga-pixels in resolution.
Digitization opens up the potential for more effective storage, optimized and standardized visualization and transmission (tele-pathology).
However, to outweigh additional effort and thereby make digitizing in pathology attractive for the routine utilization, tools for computer aided analysis are indispensable.
Automated methods can provide support by facilitating basic routine tasks, such as counting objects or segmenting regions.
Moreover, state-of-the-art machine learning approaches exhibit a potential for recognising patterns which cannot be easily detected, even by the trained human eye~\cite{myKooi17a}.
Therefore, especially for less experienced pathologists, machine learning approaches exhibit a high potential, not only to decrease the time needed, but also to improve the diagnostic accuracy. A further motivation is provided by a considerable inter-rater variability in histological examinations~\cite{mytsuda00, mypersson14}.

Microscopic evaluation of tissues or cytological preparations is the gold standard in clinical diagnostics in case of a large range of pathologies. Examples of these are smear tests, analysis of the boarders of cancer tissues during operations and post-mortem histological testing.
Due to an increasing prevalence and thus workload in the field of cancer-related diseases in combination with a decrease in the number of pathologists~\cite{myMetter19a,myPetriceks18a}, automated assisting tools will be of main importance in the near future.
By facilitating effective automated computer-based processing, image analysis approaches can be a powerful tool to facilitate clinical practice. Apart from supporting the pathologists' daily routine, automated high-throughput processing techniques can be employed to boost the potential of histological research regarding medical and biological data.

\subsection{Image Analysis in Digital Pathology}
A particularly relevant application field of digital pathology is given by the detection of tissue-of-interest combined with a pixel-accurate segmentation. Tasks such as nucleus~\cite{myMahmood19a,myChanhoJung10a,myBug19a}, cancer~\cite{Xu14a,myJiang18a} and gland~\cite{mySirinukunwattana17a,myBenTaieb16a} segmentation have been considered in recent studies. Segmentation approaches combined with the extraction of features, such as quantity, area and morphological characteristics, allow for access to image information in an efficient and condensed manner.
Classification approaches~\cite{myHou16a,myGecer18a} go one step further and have the potential to provide an observer-independent decision. While such approaches are completely automated and observer independent, an open issue in practice is how to deal with these black-box decisions when the estimated performance measure (e.g. F-score) does not indicate a perfect categorization (even if an algorithm is as accurate as a human expert). 
Stain normalization~\cite{myKhan14a,myMacenko09a,myReinhard01a} also exhibits an important field, allowing harmonization of data from a single or from several different laboratories showing stain variability. Moreover, normalization can be used for pre-processing computer-based analysis methods as well as to enhance manual experts' examination performance. 

From a technical point of view, a wide range of different approaches have been applied to histological image data.
Before the era of deep learning, for the purpose of segmentation particularly pipelines based on thresholding~\cite{myKowal19a}, watershed~\cite{myAbdolhoseini19a}, active contours~\cite{myMosaliganti08a} and a combination of these approaches were proposed. 
Stain normalization approaches were mainly based on pixel-level transformation~\cite{myKhan14a,myMacenko09a,myReinhard01a}, such as color deconvolution~\cite{myKhan14a}.
Pixel-level in this context means that mappings are generated without incorporating the pixel neighborhood.
Classification approaches were based on separate feature extraction (e.g. Local Binary Patterns~\cite{myOjala02a}, Fisher Vectors~\cite{mySanchez13a}) and classification models such as Support Vector Machines~\cite{Gadermayr16e}.

Recently, deep learning approaches and particularly Convolutional Neural Networks (CNNs) where identified as highly powerful and generic tools, being capable of performing a large range of tasks~\cite{myDimitriou19a,myLitjens17a}.
In many application scenarios, deep learning methods outperformed the existing approaches~\cite{myDimitriou19a}. Especially in the field of segmentation, the so-called fully-convolutional networks using skip-connection~\cite{myRonneberger15a,myBenTaieb16a,Gadermayr18e}, such as the prominent U-Net~\cite{myRonneberger15a}, boosted segmentation accuracy and exhibited high efficiency. Thereby allowing for a rapid processing of huge images in combination with rather cheap consumer graphics processing units.

\subsection{Challenges}
A disadvantage, however, of many deep learning approaches is given by the fact that these methods typically need large amounts of labeled training data. Data augmentation can be a powerful tool to lessen this restriction~\cite{myWei19a,myZhao19a,myRonneberger15a}. Nevertheless, a significant amount of manually annotated data is mostly indispensable. Due to the large image size of up to several gigapixels, manual annotation of histological whole slide images for the purpose of segmentation can be extremely time-consuming. As this task often needs to be performed by medical experts, this fact exhibits a burden for the application of deep learning approaches in practice. A further difficulty arises due to the variability in the image domain~\cite{myReinhard01a} which is typically (unintentionally) caused by differences in the cutting and staining process. Intentional differences can also be due to other staining techniques, applied in order to extract other or additional features from the image data. Varying staining techniques showing similar morphologies, but different texture and color characteristics, also require individually trained image analysis models. 
A further source for variation is given by intra-subject variability, e.g. due to (a wide range of) different pathologies.
If a standard deep learning pipeline (without domain adaptation) is applied, this urges for manually annotated training data, covering the whole range of image characteristics, which can be extremely diverse if several degrees of variation occur~\cite{Gadermayr19b}. 

\subsection{Contribution}
Approaches relying on Generative Adversarial Networks~\cite{myGoodfellow14a}~(GANs) exhibit the potential to reduce the requirement of large amounts of manual annotations. This, in turn, reduces the barrier to entry for automated image analysis methods in medical imaging. Particularly in the field of digital pathology, recent developments not only improved measures but even enabled novel applications. Many tasks for which supervised learning approaches were indispensable,  can now be performed with unsupervised techniques.

In this paper, we summarize the application scenarios and recent developments including GAN-based approaches in the field of digital pathology.
Based on this research, we highlight application scenarios which highly profit from recent GAN approaches using some of the most prominent architectures and adaptations of these architectures.
We also identify remaining issues and challenges and determine relevant high potential fields of research for the future.
Finally, we also provide uniform definitions to facilitate an orientation in the "jungle of GANs".

In Sect.~\ref{sec:architectures}, a summary and classification (based on capabilities) of architectures applied to digital pathology are provided.
In Sect.~\ref{sec:tasks}, the histological application scenarios are outlined followed by a review of the individual approaches (Sect.~\ref{sec:stainnorm}--\ref{sec:datagenaug}). In Sect.~\ref{sec:discussion}, we discuss trends, benefits, challenges and additional potential of GANs. Section~\ref{sec:conclusion} concludes this paper.

\section{GAN Architectures} \label{sec:architectures}
\begin{figure*}[htbp]
  \centering
  \includegraphics[width=0.8\textwidth]{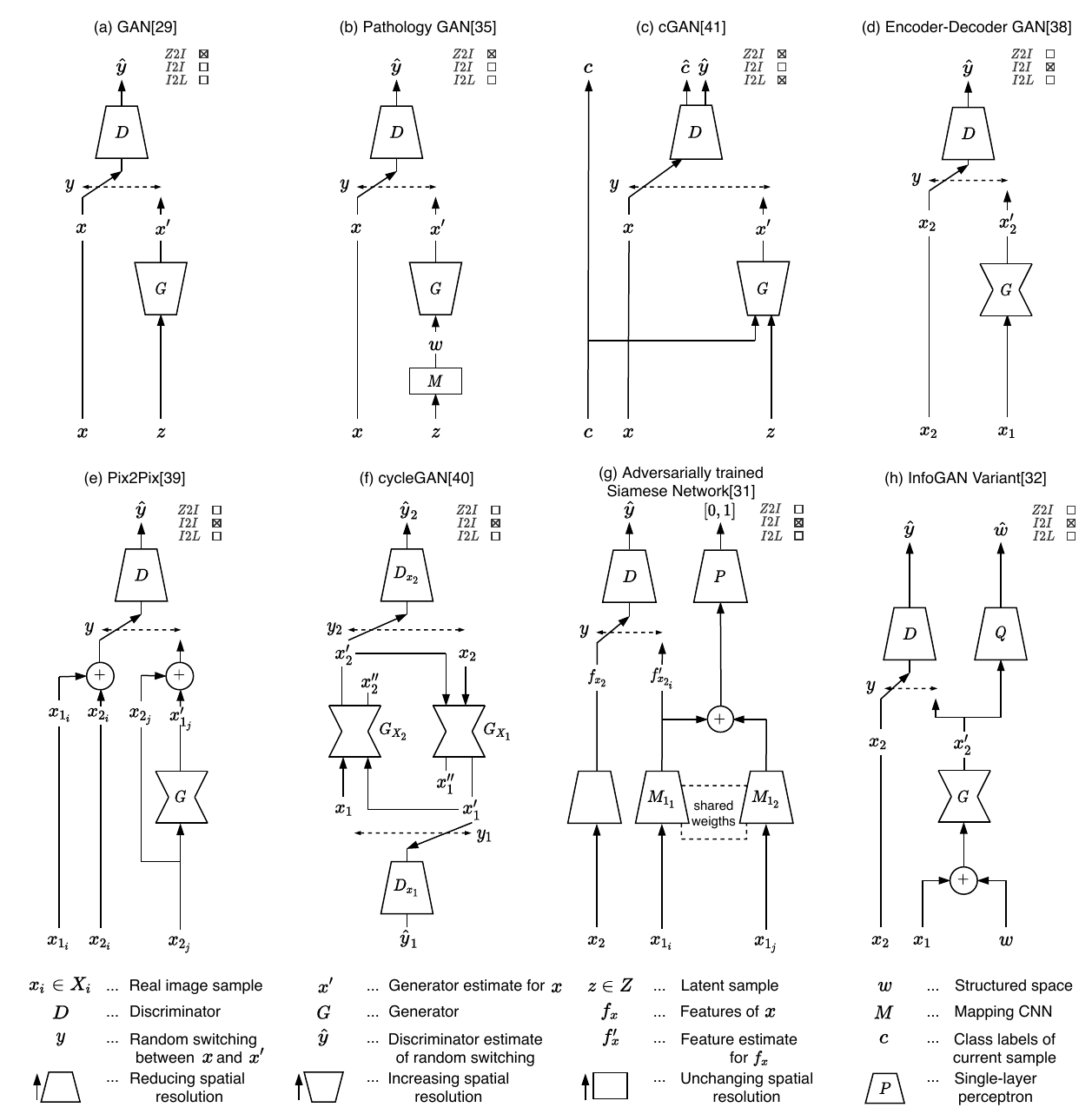}
  \caption{Architecture comparison of several Latent-to-Image, Image-to-Image and Image-to-Label networks for digital pathology, trained primarily through an adversarial loss. The form of the trapezoids describes if dimensionality is increased or reduced.}
  \label{fig:architectureCompare}
\end{figure*}

The idea of training two neural networks in a zero-sum min-max game is shown to enable stable training in image analysis for digital pathology with several important architectures like GAN, cGAN, cycleGAN, InfoGAN, BigGAN and GAN based siamese Networks \cite{myHou19a, myRen18a, myZanjani18a, myBug19a, myShaban19a, myDeBel19a, myQuiros19a, myHuo17a, myChang18a}. In this section, we focus on the technical background of GAN approaches employed for image analysis in digital pathology. We analyze these architectures and cluster them into similar groups (Fig.~\ref{fig:architectureCompare}) with respect to their applicability. Additionally, we summarize the capabilities of individual GAN architectures in Sect.~\ref{sec:cap}.

The conventional GAN architecture, introduced by Goodfellow et al. \cite{myGoodfellow14a}, is shown in Fig.~\ref{fig:architectureCompare} \textit{a}. It enables the generation of image data by mapping an unstructured latent space into an image $Z \rightarrow X$, using an up-scaling CNN called a generator $G$.
In order to generate images showing desired characteristics, this generator is trained with the aim of fooling a discriminator ($D$). The discriminator, typically also a CNN, is trained to distinguish between real ($x \in X$)  and generated samples ($G(z)$). The labeling $y$ describes the data to be real if $y = 1$ and generated if $y = 0$. Therefore, $\hat{y}$ describes the discriminator's prediction which is used by the so-called adversarial loss ($loss(y, \hat{y})$) \cite{myGoodfellow14a}. This loss is incorporated in all of the following GAN architectures.

This basic architecture can be adjusted in multiple ways and for a multitude of tasks. These adjustments are e.g. the addition of a mapping layer before the generator (Fig.~\ref{fig:architectureCompare} \textit{b} \cite{myQuiros19a}) or the replacement of the generator with an encoder-decoder structure (Fig.~\ref{fig:architectureCompare} \textit{d} \cite{myBentaieb18a}). By replacing the generator with an encoder-decoder structure (Fig.~\ref{fig:architectureCompare} \textit{d}), image transformations from one domain to another are enabled \cite{myIsola16a}. This can be achieved by replacing the latent space $Z$ with images of the source domain $X_1$ and training the generator to transform $X_1$ into a realistic image of the domain $X_2$ using an adversarial loss. This enables applications like e.g. segmentation ($X_1 \rightarrow X_2$). Adding a mapping network between $Z$ and the generator (Fig.~\ref{fig:architectureCompare} \textit{b}) aims to map the unknown latent space $Z$ to a structured latent space $W$. This interpretable structure enables semantic vector operations that translate into domain specific feature transformations \cite{myQuiros19a}. 

CGAN (Fig.~\ref{fig:architectureCompare} \textit{c}), in comparison to the original GAN approach (Fig.~\ref{fig:architectureCompare} \textit{a}), adds class information ($c$) to both the real ($x$) and generated samples ($x'$). Therefore, the discriminator checks if $G(z|c)$ and $x$ is real or generated and from the correct class. This architecture is capable of constructing images from different classes based on a single generator~\cite{mymirza14}.

Pix2Pix~\cite{myIsola16a} (Fig.~\ref{fig:architectureCompare} \textit{e}) is a variation of cGAN which replaces the up-scaling generator for an encoder-decoder structure and the class information with the corresponding image from the second domain. Therefore, the Pix2Pix generator learns to translate between two image domains ($X_{1} \rightarrow X_{2}$). 
For training, Pix2Pix needs corresponding samples (pairs) showing images from the two domains capturing the same underlying content.
Therefore, the requirements are the same as for (fully convolutional) segmentation networks, such as U-Net~\cite{myRonneberger15a}.

CycleGAN~\cite{myZhu17a} is a type of GAN (Fig.~\ref{fig:architectureCompare} \textit{f}) that enables unpaired training of image translation through an adversarial loss in combination with a cycle-consistency loss. The core idea is to train two generators to transfer images from domain $X_1$ to domain $X_2$ and vice versa.
Therefore, the loss can be calculated by combining the adversarial loss ($loss(y_1, \hat{y_1})$) with a cycle-consistency loss ($loss(x_1,x_1'')$ with $x_1''$ being $G_{X_1}(G_{X_2}(x_1)$) for both image domains. 
This cycle consistency loss penalizes changes in structural information from the real to the reconstructed sample \cite{myZhu17a}.
Without a further constraint, the generators typically also maintain the structure in the virtual domain. This is probably because a significant modification of the underlying structure followed by the inverse modification would be more complex to learn.

Combining the idea of adversarial learning with a siamese network ~\cite{koch15}, as shown in Fig.~\ref{fig:architectureCompare}~\textit{g}, results in a feature based domain transfer method \cite{myRen18a}. The leftmost part of this network structure ($M_2$) is a CNN that is trained in a supervised manner and encodes the data from domain $X_2$ in a feature space $f_{x_2}$. The CNN $M_{1_1}$ is trained adversarially, in order to extract similar features from $X_1$ and $X_2$. 
In order to keep the domain information of $X_1$, $M_{1_1}$ and $M_{1_2}$ are trained as a siamese network. As shown by Fig.~\ref{fig:architectureCompare}, the features from $M_{1_1}$ and $M_{1_2}$ are concatenated and evaluated by a CNN on how well the domain information is kept. These networks share weights and finally $M_{1_1}$ generates the features for an unseen sample of domain $X_1$ in the domain $X_2$.

The InfoGAN variant~\cite{myChen16} introduced in Fig.~\ref{fig:architectureCompare}~\textit{h}~\cite{myZanjani18a} shows a variation on the idea of PathologyGAN (Fig.~\ref{fig:architectureCompare} \textit{b}) that aims to add structure to the latent space in order to have control over the generator results and the kind of features it produces. The InfoGAN~\cite{myZanjani18a} variation aims to find a structured latent space ($W$) to decouple the color information from the underlying image information. This is accomplished by first concatenating the noise matrix with the image sample from the second domain, and then applying the result into an encoder-decoder structured generator. This generator is trained adversarially as well as through the mutual information of the auxiliary network that aims at separating the reconstruction from its structured latent part.

\subsection{GAN Capabilities in Image Analysis}\label{sec:cap}
For differentiation of the capabilities, we decided to use a generic scheme (as indicated in Fig.~\ref{fig:architectureCompare}). We defined the applications as mappings from one input domain to another output domain. Particularly, we differentiated between Latent-to-Image (Z2I), Image-to-Image (I2I) and Image-to-Label (I2L) translation.

\paragraph{Latent-to-Image (Z2I)} The application corresponds to the original GAN's~\cite{myGoodfellow14a} idea of generating images out of noise. This results in a network that can produce a theoretically infinite number of images based on unstructured latent samples ($z \in Z$).
Such a mapping ($Z2I: Z \rightarrow X$) is typically performed by conventional GANs, cGANs and their various modifications such as the progressive-growing GAN \cite{levine20, karras17} and Wasserstein GAN \cite{myarjovsky17}, partially displayed in Fig.~\ref{fig:architectureCompare} \textit{a}, \textit{b}, \textit{c} and \textit{h}. Additionally, latent samples can be mapped to a structured space before image generation in order to enable interpretable modifications \cite{myQuiros19a ,myZanjani18a}.

\paragraph{Image-to-Image (I2I)}
Another typical application of GANs can be summarized as Image-to-Image translation. 
I.e. a mapping from one image domain $X_1$ to another image domain $X_2$ is learned ($I2I: X_1 \rightarrow X_2$).
For means of generalization, we explicitly also categorized segmentation mask domains as image domains.
We decided on this generalization since the same technical approaches are used for the purpose of Image-to-Image, Image-to-Mask (known as segmentation) and Mask-to-Image translation (known as image synthesis).
For this purpose, Pix2Pix, CycleGAN (Fig.~\ref{fig:architectureCompare} \textit{e} and \textit{f}) as well as further GANs like the encoder-decoder GAN (Fig.~\ref{fig:architectureCompare} \textit{d}) and the adversarially trained siamese networks (Fig.~\ref{fig:architectureCompare} \textit{g}) can be applied.
Training of I2I approaches can be categorized into two major classes, namely paired (Pix2Pix and InfoGAN) and unpaired (cycleGAN, encoder-decoder GAN and the adversarial siamese network). While paired training requires corresponding samples from the two domains for training, unpaired training only needs two individual data sets from both domains.
While paired approaches typically exhibit better performance, unpaired techniques enable additional areas of application since paired data is not always available ~\cite{myZhu17a}.

\paragraph{Image-to-Label (I2L)}
Image-to-Label translation is typically referred to as classification. A network is trained to find a mapping $I2L: X \rightarrow \{0,1, ..., n\}$ from an image domain $X$ to a label domain comprising $n$ classes.
For this purpose, cGANs and cGAN variants can be employed (Fig.~\ref{fig:architectureCompare} \textit{c}), since the discriminator is, apart from a discrimination between real and fake samples, also trained to determine the class label~\cite{lecouat18}.


\section{Tasks in Digital Pathology}\label{sec:tasks}

I2I, I2L and Z2I correspond to a multitude of applications in histological image analysis.
Here, we identified applications specific for digital pathology and assigned them to one of the three translation settings.

Particularly, we identified stain normalization (Sect.~\ref{sec:stainnorm}), stain adaptation (Sect.~\ref{sec:staindomainadapt}), segmentation based on supervised learning (Sect.~\ref{sec:supervisedseg}), the synthesis of image data based on segmentation masks (Sect.~\ref{sec:syn}) and data augmentation as translation settings (Sect.~\ref{sec:datagenaug}).
Subsections \ref{sec:stainnorm}--\ref{sec:datagenaug} can be categorized as I2I.
Data augmentation (Sect.~\ref{sec:datagenaug}) can be both, I2I and Z2I, depending on the specific configuration.

We defined stain normalization (Sect.~\ref{sec:stainnorm}) as a mapping from an original image domain $X_o$ to a normalized domain $X_n$. $X_n$ can be a subset of $X_o$, necessarily showing lower variability.
Apart from the stain, the mapping is intended to keep all further image characteristics.
With stain adaptation (Sect.~\ref{sec:staindomainadapt}), we refer to the setting where not (only) the variability within one staining protocol (e.g. H\&E), but between different protocols, need to be compensated. Therefore, the domain shift is supposed to be larger than in case of stain normalization. In this subsection, we also include a general domain adaptation, which is not a typical I2I setting (as the adaptation is typically not performed on an image but on a feature level). Anyway, we decided for this categorization due to the similarity from the application's point of view. According to our definition, stain adaptation can be interpreted as a special type of domain adaptation.

Z2I in digital pathology is utilized in order to increase the data set size.
In this paper, data augmentation refers to the I2I setting and data generation refers to the Z2I configuration (Sect.~\ref{sec:datagenaug}).
This, in combination with additional transformations, like converting the unstructured into structured latent spaces, allows e.g. for the generation of a theoretically infinite number of morphologically different images.

I2L is mainly used for final classification in diverse applications (Sect.~\ref{sec:stainnorm}--\ref{sec:datagenaug}). Theoretically, the discriminator of a cGAN and cGAN-like networks can typically be re-purposed for I2L \cite{mymirza14} but only few applications use the cGAN discriminator directly \cite{myburlingame18}. Additionally, the discriminator of most GAN variations can be used for representation learning, which can in turn be used as input of subsequent evaluation through e.g. SVMs \cite{myhu18}.

\subsection{Stain Normalization} \label{sec:stainnorm}

Since stain normalization is a type of I2I translation, several GAN based approaches, as introduced in Sect. 2, can theoretically be used to enable stain normalization. CycleGAN (Fig.~\ref{fig:architectureCompare} \textit{f}) can be optimized for the means of stain normalization based on one training data set from a general ($X_o$) and a normalized domain ($X_n$). Pairs, which are hard to collect for this scenario, are therefore not needed. De Bel et al. \cite{myDeBel19a} investigated various experimental settings with different generator architectures combined with data augmentation strategies for cycleGAN. They showed that stain normalization using the baseline architecture performs well and eliminates the need for any further stain augmentation. It is shown in general, that cycleGAN is highly flexible and powerful and exhibits a general purpose architecture which is capable of stain normalization. Compared to common pixel-based stain-normalization approaches~\cite{myKhan14a,myMacenko09a,myReinhard01a}, however, cycleGAN can do more than applying a non-linear pixel-based mapping. The approach is theoretically also able to generate changes in texture. Depending on the used data sets for training, this capability corresponds to the potential of introducing bias. This is especially the case if $X_o$ and $X_n$ show systematic differences regarding the underlying tissue characteristics (e.g. in case of data showing variable degrees of pathologies). Experiments however, prove good performance in general, also with respect to final segmentation or classification tasks~\cite{myDeBel19a}. 
However, an experimental investigation of the impact of different distributions in the two data sets used for training the model has not been performed so far.
To eliminate bias in these kinds of architectures, the stain normalization stage can be integrated into a classification approach~\cite{myBentaieb18a}.

Other approaches performing unpaired Image-to-Image translation for stain normalization, replace the cycle consistency loss with an alternative formulation. In the following, we summarize two such approaches. 
The authors of \cite{myBentaieb18a} used an encoder-decoder GAN (Fig.~\ref{fig:architectureCompare} \textit{d}) with an additional loss to keep morphological consistency. This is achieved through a further gradient loss. Additionally, this stain normalization model is combined with a classification model, which can be used to add a classification loss in order to optimize the separability of the classes. The potential of the additional loss here faces reduced flexibility, since the model needs to be trained individually for each task. Anyway, the limitation is modest as classification models are necessarily trained or adapted for each specific task.
%

A different stain normalization approach is based on the idea of InfoGAN (Fig.~\ref{fig:architectureCompare} \textit{h}). The authors of~\cite{myZanjani18a} replaced the latent space $Z$ with the lightness channel of the source image. Additionally, a mutual information loss is used to train the generator to represent the structured space $W$ as the color transformation of the source image. This allows normalization into a predefined structured space or in the case of [32], a predefined color space.





\subsection{Stain and Domain Adaptation} \label{sec:staindomainadapt}

One approach to perform domain as well as stain adaptation is by the use of cycleGAN (Fig.~\ref{fig:architectureCompare} \textit{f}) as introduced in~\cite{Gadermayr18d}. The authors performed segmentation based on previously adapted image data. While the adaptation method is based on cycleGAN and very similar to stain normalization \cite{myDeBel19a,myShaban19a}, focus here is on translating one histological stain into another. For the purpose of domain adaptation, the images are virtually re-stained before processing. This approach is based on the assumption that annotated training data is only available for one certain stain which is approached by translating each stain to the target stain.

Similarly, Xu et al.~\cite{myXu19a} adapted cycleGAN to translate between different stains (H\&E and immunohistochemistry (IHC)) through the addition of a structured loss, aimed at suppressing bias (here referred to as "imaginary" features). Contrary to~\cite{Gadermayr18d}, the authors did not apply further image analysis, such as classification or segmentation, but instead evaluated the realism of the data through Turing tests. The authors identified a large set of application fields including a fast and low-cost generation of IHC stainings, virtual multiplexing, co-localization, augmentation and color deconvolution.

In a similar manner, an approach was proposed to translate between H\&E and immunofluorescent stains~\cite{myChang18a}. Unlike the approaches mentioned before~\cite{Gadermayr18d,myXu19a} based on unpaired training (cycleGAN), here Pix2Pix (Fig.~\ref{fig:architectureCompare} \textit{e}) is employed. Typically, paired samples showing exactly the same tissue in two different stains, are hard to collect. 
The authors here, however, used multiplexed imaging which allows for the generation of perfectly corresponding image pairs.
Even though multiplexing could also be applied clinically, it is costly and can degrade both tissue quality and antigenicity. This kind of I2I translation might omit the need for clinical multiplexing.

Another Pix2Pix approach by Rana et al.~\cite{myRana18a} similarly utilized I2I translation based on paired training to convert H\&E to unstained image data (and vice versa). Pairs are available, since the unstained sample is captured before applying the H\&E staining. Only additional registration is needed to obtain the pixel correspondences. Both approaches~\cite{myRana18a,myChang18a} qualitatively seem to be able to successfully use Pix2Pix as an I2I translation model, but since they lack an evaluation in the form of subsequent classification or segmentation, a quantitative conclusion cannot be drawn. 

Additionally, the siamese GAN architecture introduced in ~\cite{myRen18a} (Fig.~\ref{fig:architectureCompare} \textit{g}) can be used for domain adaptation. This domain adaptation pipeline maps the source domain to the target domain in an unsupervised, feature based (not image based) manner. By combining adversarial and siamese training procedures, features from the source domain are mapped to the target domain, while still being kept structurally similar to the source domain. Finally, The obtained results are successfully evaluated on the classification of image patches.


\subsection{Supervised Segmentation} \label{sec:supervisedseg}

The Pix2Pix network is an established powerful segmentation approach, exhibiting an alternative to conventional fully-convolutional segmentation networks~\cite{myRonneberger15a}.
Wang et al.~\cite{myWang17a} adapted the Pix2Pix model (Fig.~\ref{fig:architectureCompare} \textit{e}) for the field of histology and showed improved performance compared to a standalone fully-convolutional network~\cite{myYang16b}. For many tasks, standalone fully-convolutional networks~\cite{mySirinukunwattana17a,myRonneberger15a,myBenTaieb16a,Gadermayr17d} (without an adversarial loss) perform reasonably well in histopathological image analysis. However, a difficulty here is given by the fine structure of the basal membrane. The GAN's advantage over these basic (non-adversarial) CNN approaches is given by the ability to maintain fine details through the adversarial loss~\cite{myIsola16a}.

Further potential of GANs as a principally basic segmentation scenario is shown in~\cite{Gupta19a} where
GANs were applied in order to "enrich" the image domain in a segmentation setting.
Instead of increasing the number of samples, as performed in case of data augmentation, the information per image was enlarged.
By performing image translation between a source and several target stains for each image, additional virtually re-stained images were created.
This generated data was used to train and test a segmentation network which outperformed the baseline setting of a network trained and tested on the source domain only. However, a positive effect of this application was only shown in one very specific application in the field of kidney pathology.


\subsection{Synthesis enabling Unsupervised Learning} \label{sec:syn}

One way to synthesise histological samples starts with the generation of label masks.
Realistic shapes are obtained by sampling the parameters of the objects randomly from distributions corresponding to natural occurrences.
Mahmood et al.~\cite{myMahmood19a} directly used binary masks and added realism by deploying a generator of a cycle-GAN model. This model was trained on an unpaired data set consisting of a label-domain and an image-domain data set. 
Bug et al.~\cite{myBug19a} proposed a similar approach. Instead of converting binary masks and images, the authors already added a certain kind of realism before GAN-based image translation. Specifically, they added typical colors as well as spot-noise combined with blur in the back- as well as foreground.
Hou et al.~\cite{myHou19a,myHou17a} put even more focus on hand-crafted synthesis. Background was obtained from original images by making use of unsupervised segmentation techniques. In a similar way, object-texture was obtained from real objects.
Back- and foreground were then combined and an encoder-decoder GAN (Fig.~\ref{fig:architectureCompare} \textit{d}), was trained to translate from the pre-computed synthetic domain to the real domain. Instead of the cycleGAN architecture, this encoder-decoder GAN with one generator and one discriminator is utilized, containing a regularization loss ($L_1$ and $L_2$ norm between the input and output image), a discriminator loss and a task-specific loss~\cite{myShrivastava16a,myShrivastava17a}, to focus on the generation of challenging samples.

Senaras et al~\cite{mySenaras18a} utilized a similar architecture (based on Pix2Pix) to generate realistic histopathological images from ground truth label-mask images. The method is similar to~\cite{myHou19a,myHou17a} using an $L_1$ and a discriminator loss. Compared to the other approaches~\cite{myMahmood19a,myBug19a,myHou17a,myHou19a}, the goal here is not the (unsupervised) segmentation, but the generation of a data set to be used for analysis of computer-based algorithms as well as inter- and intra-observer variability. Instead of artificially generated label masks, the authors translated ground-truth annotations into realistic images. The obtained pairs where intended to show perfect correspondence, which is not the case if data is annotated manually in the traditional sense.

A vice versa approach (compared to \cite{myMahmood19a,myBug19a,myHou17a,myHou19a}) was proposed in~\cite{Gadermayr19c, Gadermayr19b}. Instead of generating virtual images out of label-masks for means of obtaining labeled training samples, the authors performed translation directly from the image to the label-mask domain. Similar to~\cite{myMahmood19a}, a cycleGAN model was trained to translate from the label-mask to the image domain and vice versa. Ultimately, the authors used the generator translating images to label and thereby immediately obtain segmentation output, circumventing the need for an additional segmentation model.


This translation between the image and the segmentation label-mask domain (and vice versa) exhibits a highly interesting field, with the potential for unsupervised segmentation. The unpaired and thereby unsupervised approach can be combined with manually labeled data to improve performance even further (dependent on the amount of labeling resources). Based on the publications so far, it is hard to make a general statement, whether a translation from the label-mask to the image domain or vice versa is more effective with regards to unsupervised segmentation. Performing a translation from the image to the label-mask domain, is probably a task which is easier to learn. The authors of~\cite{Gadermayr19c, Gadermayr19b} were unable to generate realistic images, whereas the translation from the image to the label-mask domain showed reasonable outcomes.

The limitation is given because the mapping from the image to the label-mask is more or less defined, while the vice versa mapping is ambiguous (also referred to as ill-posed). This means that for one label-mask, there exist several corresponding images. In case of cycleGAN, both mappings (i.e. both generators) need to be trained, independent of the finally needed generator. The "ambiguous" mapping, however, can affect training based on the cycle-consistency loss, as the loss $||x_1 - G_{X_1} \circ G_{X_2} (x_1)||_2$ is not necessarily small, even if the generators show attended behavior. If the mapping represented by the generator $G_{X_1}$ is ambiguous, a variety of realistic images can be generated out of $G_{X_2}$. This issue is discussed in detail in~\cite{Gadermayr20a}.

Another approach towards unsupervised or weakly supervised learning is given by representation learning. 
Hu et al.~\cite{myHu19a} adapted the GAN architecture for learning cell-level image representations in an unsupervised manner. For that purpose, an auxiliary network was employed, which shares weights with the discriminator. In addition to the discriminator loss, the authors introduced a further mutual information loss. The trained auxiliary network can be employed to extract features on cell-level, which are used to perform cluster analysis. The authors utilized the aggregated cluster information to train an image-level classification model. However, these extracted features could also be applied for high-level image segmentation.


\subsection{Data Generation \& Augmentation} \label{sec:datagenaug}

Similar to work on domain adaptation~\cite{Gadermayr18d}, Wei et al.~\cite{myWei19a} adapted the cycleGAN architecture in order to perform data augmentation. Instead of performing what we typically refer to as domain adaptation (i.e. the adjustment between slightly dissimilar distributions while the class labels remain similar), they trained the GAN architecture in order to translate from one tissue category to another (here from normal to abnormal). Thereby, they obtained a generation model for the means of data augmentation which creates, based on existing samples, additional samples of the other class.
The difficulty of this task is, that not only low-level image details, such as color, need to be changed. In contrast, a translation from one class to another typically requires a major change of the image morphology. The authors showed that this is actually effective in the considered application scenario~\cite{myWei19a} as the achieved classification performance could be increased. However, the cycleGAN architecture in general is not optimized for performing morphological changes. Similarly, as in Sect.~\ref{sec:syn}, the problem of ambiguous mapping emerges. The problem in case of synthesis is, that one mapping (from the label to the image domain) is ambiguous. Here, both mappings are potentially ambiguous, because there is typically not a one-to-many, and definitely not a one-to-one, mapping corresponding to pathological changes.

Another approach for data generation was proposed in ~\cite{myQuiros19a}. The focus was on generating artificial cancer tissue from a structured latent space (Fig.~\ref{fig:architectureCompare} \textit{b}). This method was not evaluated on any following classification or segmentation. Instead, its quality was bench-marked based on the Fr\'enchet Inception Distance.

\section{Discussion: Potential of GANs}\label{sec:discussion}
We identified three fields in digital pathology with a particularly high potential of GANs.
These fields were identified based on the related work summarized in this paper in Sect.~\ref{sec:tasks}. 
Certainly, we do not claim the exclusive truth. In contrast, we invite the reader to have a critical look at this review and expand on it in future research.

\subsection{Synthesis instead of Labeling}
Firstly, we assessed the capability of cycleGAN and derivatives to translate from an image to a label domain (and also vice versa) as an extremely powerful approach. The ability to learn from unpaired data in this settings translates the labeling problem into a synthesizing problem. For many applications in the field of digital pathology, synthesis of realistic ground-truth label maps is a feasible task. Particularly, as long as roundish-shaped objects need to be created, a basic simulation model relying on a handful of parameters is sufficient~\cite{myMahmood19a,Gadermayr19c,myBug19a}.

A difficulty here is given by the fact that the mapping from the label to the image domain is mostly ambiguous, as a label-mask can be mapped to more than one corresponding image.
This potentially complicates training of the complete GAN architecture with diverse methods of resolution.

One approach to tackle this challenge is to change the cycleGAN architecture, e.g. by skipping one of the two cycle-consistency loss terms~\cite{Gadermayr20a}. The idea here is that ambiguous mappings affect the loss in one of the two cycle-consistency mappings only. Alternatively, Almahairi et al.~\cite{myAlmahairi18a} and Huang et al.~\cite{myHuang18a} proposed a cycle-GAN extension employing auxiliary latent spaces to control the variations of the one-to-many mappings. The idea is, to decompose an image into a content code that is domain-invariant and a domain code which captures domain-specific properties.

Another method of resolution is outlined in~\cite{Gadermayr19b}. The authors showed that cycleGAN training is more effective when the label domain contains additional information. With additional information, the authors do not mean noise, but information which is also available in the image domain. In addition to the roundish high-level objects-of-interest, they also synthesized low-level information (nuclei) and showed that thereby the performance and robustness can be increased dramatically. The improvement is obtained, because the generator networks now receive information where to place the low-level objects in order to obtain a low cycle-consistency loss.
The difficulty here is, that the simulation model thereby becomes more complex. However, especially in case of higher-level objects, we are confident that this is the most optimal solution when unpaired approaches, such as cycleGAN, should be trained to perform translations from a label-mask to an image domain or vice versa.

For practical utilization, an adjustable simulation tool would be highly helpful to generate data according to the characteristics of an individual data set. As segmentation tasks in digital pathology often correspond to rather uniform roundish objects, basic functionality would be enough for many purposes.

So far, such approaches have not been applied and adapted to applications without strong shape constrains, as in tumor segmentation~\cite{myHou16a}. Apart from the diverse morphology, a difficulty here is given by the scale of the regions-of-interest. Regions can show up to several hundreds or even thousands of pixels in diameter. As the segmentation networks are applied patch-wise, this can constitute a severe challenge. A method of resolution might be given by a multi-level (or multi-resolution) approach. Small, morphologically regular structures (such as nuclei) could be extracted in a first step. Afterwards, a segmentation of high-level objects (such as stroma, tumor, etc.) can be performed on a lower resolution, based on the segmentation information extracted on the lower level.

\subsection{Potential of Stain-to-Stain Translation}

We further identified stain-to-stain translation as an application with high potential.
In previous work, stain-to-stain translation was performed mainly for means of domain adaptation. However, previous work already showed, that a GAN is capable of facilitating the segmentation task by either changing the appearance or by adding additional information to the image domain~\cite{Gupta19a}.

When we think of domain adaptation, we typically do not focus on facilitating a segmentation problem, but aim at adjusting one data set to another. Nevertheless, we are confident that a previous conversion can be a helpful tool to improve segmentation or classification analysis performance. For this purpose, special stains can be used, which particularly highlight the respective objects-of-interest without the need for physical generation of additional slides. 
Another option is given by the translation from light field to fluorescence microscopy~\cite{myChang18a}. Which, in addition, has the potential of trivializing the following segmentation task. 

A question which has not been addressed so far is, whether a translation from a stain to another (e.g. a general purpose stain, such as H\&E, to an IHC stain) is capable of showing similar features as the real target stain. In~\cite{Gadermayr18d}, such a translation was performed, but only for segmenting higher level objects which are independent of the histological stain. The requirement here was only that the morphology of the (high level) objects-of-interest are maintained.
To show whether or not GANs are able to generate virtual stains which are not only realistic, but also exhibit similar features on a low-level (i.e. the stain response on pixel-level) as a real stain, a special data set is needed. Particularly, this requires a large set containing corresponding slide pairs stained with two different approaches. Apart from the capabilities of the neural networks, the principal question here is whether the information is available in the image data or not, which might differ from problem to problem. Anyway, due to the immense impact, we are confident that this is worth trying out in a defined application scenario.

\subsection{Morphology Translation}

Image-to-Image translation typically covers mappings from one domain to another, where the domain gap is caused by (intentional or unintentional) variability in the data generation protocol~\cite{myShaban19a,myZanjani18a,myChang18a,myRana18a,myXu19a,Gadermayr18d}. 
In each of these settings, color and potentially also texture varies between the domains. However, as the underlying tissue is similar, there are mostly no clear morphological changes.

A setting with morphological changes has been investigated in~\cite{myWei19a}. The authors of this paper explored a translation between data from different classes for the means of data augmentation. They used a derivation of the cycleGAN architecture and achieved improvements regarding the final classification task, but also figuring out that there is a clear difference between the generated and the real image data. This statement is reinforced by the finally obtained classification rates, which were higher for real than for virtual data.

Even though cycleGAN generally shows a high versatility, we are confident that it does not exhibit the optimum architecture for settings with changed morphology. In case of morphological changes, there occur typically ambiguous mappings which can be highly problematic (as already discussed in Sect.~\ref{sec:datagenaug}).

Furthermore, conventional CNN architectures are not optimally suited to perform spatial translation~\cite{myJaderberg15a}. Subsequently, we are confident that this field exhibits potential for further improvements with specifically optimized architectures. Methods including a spatial transformer module~\cite{myJaderberg15a} might be beneficial for this purpose. A powerful tool could facilitate a translation between healthy and pathological data, e.g. for the means of data augmentation. In addition, a translation between different imaging settings (such as frozen-to-paraffin translation) might be considered, potentially improving the image quality and therefore also the final classification accuracy.

\section{Conclusion} \label{sec:conclusion}
In this paper, we summarized existing GAN architectures in the field of histological image analysis. We provided an overview of addressed application scenarios and the employed methods and identified the major fields of research.
Apart from current trends and benefits through GANs, we also identified remaining potential and the urge for novel technical approaches to improve image analysis even further. In general, it can be stated that GANs exhibit the potential for relaxation or even for elimination of the constraint that large amounts of annotated training data are needed for training deep neural network architectures. In spite of remaining challenges, we think that this technology will play a key role when it comes to practical applicability of flexible image analysis methods in digital pathology.


\bibliography{my,eigene}
\bibliographystyle{ieeetr}


\end{document}